\documentclass[12pt,a4paper,english]{iopart}
\usepackage{times}
\usepackage[T1]{fontenc}
\usepackage[latin1]{inputenc}
\usepackage{babel}
\usepackage{graphics}
\usepackage{latexsym}
\usepackage[colorlinks,ps2pdf]{hyperref}

\begin{document}

\title{\textcolor{blue}{The quark strange star in the enlarged Nambu-Jona-Lasinio model}}

\author{\textcolor{blue}{Ryszard Ma\'{n}ka}\footnote{
\textcolor{green}{manka@us.edu.pl}}, \textcolor{blue}{Ilona Bednarek}
\footnote{
\textcolor{green}{bednarek@us.edu.pl}} 
~ and \textcolor{blue}{Grzegorz Przyby\l a}}

\address{Institute of Physics, University of Silesia, Katowice 40007, ul.
Uniwersytecka 4, Poland.}

\begin{abstract}
The strange quark star is investigated within the enlarged SU(3) Nambu-Jona-Lasinio
(NJL). The stable quark star exists till maximal configutation with
\( \rho _{m}=3.1\, 10^{15}\, g/cm^{3} \) with \( M_{m}=1.61\, M_{\odot } \)
and \( R_{m}=8.74\, km \) is reached. Strange quarks appear for density above
\( \rho _{c}=9.84 \) \( g/cm^{3} \) for the quark star with radius
\( R_{c}=8.003 \) \( km \) and \( M_{c}=0.77\, M_{\odot } \). The
comparison of a quark star properties obtained in the Quark Mean Field
(QMF) approach to a neutron star model constructed within the Relativistic Mean Field
(RMF) theory  is presented.
\end{abstract}


\submitto{\NJP}

\maketitle

\section*{\textcolor{blue}{Introduction}}

At sufficiently high density, the transition to deconfined strange
quark matter is widely expected. Chiral symmetry is spontaneously
broken in the QCD vacuum. Lattice QCD simulations at nonzero temperature
\( T \) and zero baryon-chemical potential \( \mu _{B} \) indicate
that chiral symmetry is restored above a temperature \( T\sim 150\, MeV \)
\cite{latt}. The NJL \cite{NJL} model is an effective theory which
is believed to be related to QCD at low energies, when one has integrated
out the gluon fields. The NJL model might yield reasonable results
 at  the density range where confinement is no longer crucial
but chiral symmetry as a symmetry of full QCD remains to be important.
The NJL model has proved to be very successful in the description
of the spontaneous breakdown of chiral symmetry exhibited by the true
(nonperturbative) QCD vacuum. This model has been extensively used
over the past years not only  to describe hadron properties \cite{gast} (see for
reviews \cite{Vog91,Kle92}) and phase transitions in dense matter 
\cite{Asa89,Kli90a,Kle99} but to describe the quark strange stars as well
\cite{GlenBook,Weber,Hana00,Hana01,blaschke}. The detailed
properties of the quark phase in compact stars has been a topic of
recent interest \cite{glen1} (for a review see \cite{ABR01,biel}). 

Quark strange stars are astrophysical compact objects which are entirely
made of deconfined \textit{u,d,s} quark matter (\textit{strange matter})
staying in  \( \beta  \)-equilibrium. The possible existence of
strange stars is a direct consequence of the conjecture \cite{bw}
that strange matter may be the absolute ground state of strongly interacting
matter.

The three-flavor NJL model has been discussed by many authors, e.g.
\cite{Rel}. For the quark phase we follow Buballa and Oertel \cite{BO98}
in using the three-flavor version of the NJL model.

The aim of this paper is to investigate a strange quark star within
the enlarged SU(3) Nambu-Jona-Lasinio (NJL) model. The comparison of a quark
star properties obtained in the Quark Mean Field (QMF) approach to
a neutron star model constructed in framework the Relativistic Mean Field theory (RMF)  will be
made. This paper is organized as follows.\\
In Section 1 there are presented  general properties of the NJL model
in the mean field approach based on the Feynman - Bogolubov inequality
for free energy of the system. In this section the employed equation
of state (EoS) is calculated for NJL model. The EoS is used then to
determine the equilibrium configurations of the quark star in Section
2. Finally, in Section 3 the main implications of the results are
summarized.

\section*{\textcolor{blue}{Namby-Jona-Lasinio model}}

The NJL\cite{NJL} model was widely used for describing hadron properties
\cite{thom} and the chiral phase transition \cite{riske}. The enlarged
(ENJL)\cite{MCR} simplest version of the model is given by the Lagrangian:
\begin{eqnarray}
L & = & \bar{q}(i\gamma ^{\mu }\partial _{\mu }-m_{0})q+\frac{1}{2}G_{s}\, \sum _{a=0}^{8}[(\bar{q}\lambda ^{a}q)^{2}+(\bar{q}\lambda ^{a}i\gamma _{5}q)^{2}]-2K\prod _{f=\{u,d,s\}}(\bar{q}_{f}q_{f})\nonumber \\
 &  & -\frac{1}{2}G_{v}\sum _{a=0}^{8}[(\bar{q}\gamma ^{\mu }\lambda ^{a}q)(\bar{q}\gamma _{\mu }\lambda ^{a}q)+(\bar{q}\gamma ^{\mu }\gamma _{5}\lambda ^{a}q)(\bar{q}\gamma _{\mu }\gamma _{5}\lambda ^{a}q)]\label{LagNJL} \\
 &  & +i\sum _{f=1}^{2}\overline{L_{f}}\gamma ^{\mu }\partial _{\mu }L_{f}-\sum _{f=1}^{2}m_{f}\overline{L}_{f}L_{f}+B_{0}\nonumber 
\end{eqnarray}
The first term contains the free kinetic part, including the current quark
\( q_{f}=\{q_{u},q_{d},q_{s}\} \) masses \( m_{0} \) which break
explicitly the chiral symmetry of the Lagrangian and the term representing the free relativistic 
leptons \( L_{f}=\{e^{-},\, \mu ^{-}\} \).
The fermion fields are composed of quarks and leptons (electrons,
muons). Here \( q \) denotes a quark field with three flavors, \( u \),
\( d \) and \( s \), and three colors.We restrict ourselves to the
isospin SU(2) unbroken symmetric case, \( m_{0}^{u}=m_{0}^{d} \),
thus 
\begin{equation}
m_{0}=m_{0,f}\delta _{f,f'}=\left( 
\begin{array}{ccc}
m_{0,u} &  & \\
 & m_{0,d} & \\
 &  & m_{0,s}
\end{array}\right) 
\end{equation}
Generators of the u(3) algebra \( \lambda ^{a}=\{\lambda ^{0}=\sqrt{2/3}I,\, \lambda ^{i}\} \)
(where \( I \) is an identity matrix, \( \lambda ^{i} \) are Gell-Mann
matrices of the su(3) algebra) obeying \( Tr(\lambda ^{a}\lambda ^{b})=2\delta _{ab} \).
Due to this  normalization   of this algebra the coupling constants \( G_{s} \)
and \( G_{v} \) can be redefined and written as \( \bar{G}_{s}=2/3\, G_{s},\, \bar{G}_{v}=2/3\, G_{v}. \) 
\begin{table}[!h]
\begin{center}
\noindent{TABLE 1: Parameter sets of the NJL models.} 
\begin{tabular}{|c|c|c|c|c|}
\hline 
&
NJL (su(2))\cite{riske}&
NJL (I)\cite{BO98}&
NJL (II)\cite{MCR}&
ENJL \cite{MCR}\\
\hline
\hline 
\( m_{u}=m_{d} \)&
5.5 MeV &
5.5 MeV &
5.5 MeV &
3.61 MeV \\
\hline 
\( m_{s} \)&
0&
140.7 MeV&
132.9 MeV &
88.0 MeV \\
\hline 
\( \Lambda  \)&
631 MeV&
602.3 MeV&
631.4 MeV&
750.0 MeV\\
\hline 
\( \bar{G}_{s}\Lambda ^{2} \)&
2.19&
3.67&
3.67&
3.624\\
\hline 
\( K \)&
0&
12.36&
9.40&
9.40\\
\hline 
\( \bar{G}_{v}\Lambda ^{2} \)&
0&
0&
0&
3.842\\
\hline
\end{tabular}
\label{tab1}
\end{center}
\end{table}

The NJL model is non-renormalizable, thus it is not defined until
a regularization procedure has been specified. This cut-off limits
the validity of the model to momenta well below the cut-off. In most
of our calculations we will adopt the parameters presented in Table 1. 
With \( \Lambda  \), \( G_{s} \) specified above,
chiral symmetry is spontaneously broken in vacuum.

The model contains eight parameters of the standard NJL model (the
current mass \( m_{0,u} \), \( m_{0,s} \) of the light and strange
quarks, the coupling constants \( G_{s} \), determinant coupling
\( K \) and the momentum cut-off \( \Lambda  \)) and additional constant
\( G_{v} \) (\( G_{v}=x_{v}G_{s} \), \( x_{v}=1.06 \) \cite{Hana01}).
In the quark massless limit the system has a \( U(3)_{L}\times U(3)_{R} \)
chiral symmetry. The system has following  global currents: the baryon current
\begin{equation}
J_{B}^{\nu }=\frac{1}{3}\overline{q}\lambda ^{0}\gamma ^{\nu }q
\end{equation}
 and the isospin current which exists only in the asymmetric matter
\begin{equation}
J_{3}^{\nu }=\frac{1}{2}\overline{q}\gamma ^{\nu }\lambda ^{3}q\, \, ,\, \, J_{8}^{\nu }=\frac{1}{2}\overline{q}\gamma ^{\nu }\lambda ^{8}q
\end{equation}
 The conserved baryon and isospin charges are given by the 
relations: 
\[
Q_{i}=\frac{1}{3}\int d^{3}x\, q^{+}\lambda ^{i}q,\, \, \, \, i=\{0,3,8\}
\]
which are connected to commuting Cartan algebra. The physical system is defined
by the thermodynamic potential \cite{fet}
\begin{equation}
\Omega =-kTlnTr(e^{-\beta (H-\mu ^{i}Q_{i})})
\end{equation}
where H stands for the Hamiltonian. More convenient is to  use chemical
potentials connected to the quark flavor \( f \) in the such way
that \( \mu ^{i}Q_{i}=\sum _{f}\mu _{f}\bar{Q}_{f} \) 
\begin{eqnarray}
\mu _{u}=\mu ^{0}+\mu ^{3}+\frac{1}{\sqrt{3}}\mu ^{8} &  & \nonumber \label{mun} \\
\mu _{d}=\mu ^{0}-\mu ^{3}+\frac{1}{\sqrt{3}}\mu ^{8} &  & \nonumber \label{mup} \\
\mu _{s}=\mu _{s}^{0}-\frac{2}{\sqrt{3}}\mu ^{8} &  & 
\end{eqnarray}
 Quarks and electrons are in \( \beta  \)-equilibrium which can be
described as a relation among their chemical potentials 
\begin{eqnarray*}
\mu _{d} & = & \mu _{u}+\mu _{e}=\mu _{s}\\
\mu _{\mu } & = & \mu _{e}
\end{eqnarray*}
 where \( \mu _{u} \), \( \mu _{d} \), \( \mu _{s} \) and \( \mu _{e} \),
\( \mu _{\mu } \) stand for quarks and lepton chemical potentials,
respectively. These conditions means that matter is in equilibrium
with respect to the weak interactions. If the electron Fermi energy
is high enough (greater then the muon mass) in the neutron star matter
muons start to appear as a result of the following reactions 
\begin{eqnarray*}
d\rightarrow u+e^{-}+\overline{\nu }_{e} &  & \\
s\rightarrow u+\mu ^{-}+\overline{\nu }_{\mu } &  & 
\end{eqnarray*}
The neutron chemical potential is 
\[
   \mu _{n}\equiv \mu _{u}+2\mu _{d}.
\]
 In a pure quark state the star should to be charge neutral. This
gives us an additional constraint on the chemical potentials 
\begin{equation}
\frac{2}{3}n_{u}-\frac{1}{3}n_{d}-\frac{1}{3}n_{s}-n_{e}-n_{\mu }=0.
\end{equation}
where \( n_{f} \) (\( f\in u,d,s \)), \( n_{e} \) are the particle
densities of  quarks and  electrons, respectively. The EoS can
now be parameterized by only one parameter, namely the dimensionless
\( u \) quark Fermi momentum \( x \) (\( k_{F,u}=M\, x \) (\( M=939\, MeV \)
is the nucleon mass)). 

The mean field approach means that the quantum
correlations 
\[
(A-<A>)(B-<B>)=AB-A<B>-B<A>+<A><B>\sim 0
\]
may be neglected. It allows us to substitute \( AB \) by 
\[
AB\sim A<B>+B<A>-<A><B>
\]
Using this approximation the Lagrange function \( \cal L \) may be
expressed as
\begin{eqnarray}
 & \bar{L}=\bar{q}(i\gamma ^{\mu }D_{\mu }-m_{0})q+g_{s}\, \sum _{a=0}^{8}\sigma ^{a}(\bar{q}\lambda ^{a}q)-\frac{1}{2}m_{s}^{2}\sigma ^{a}\sigma ^{a}+\frac{1}{2}m_{v}^{2}V^{a}_{\mu }V^{a\mu }+\nonumber  & \\
 & -2K\sum _{f}(\bar{q}_{f}q_{f})\prod _{f'\neq f}<\bar{q}_{f'}q_{f'}>+4K\prod _{f}<\bar{q}_{f}q_{f}> & \label{lagsug} 
\end{eqnarray}
where the covariant derivative is given by 
\begin{equation}
D_{\mu }=\partial _{\mu }+\frac{1}{2}ig_{v}V_{\mu }^{a}\lambda ^{a}
\end{equation}
Here, the meson fields first appear as non-dynamical variables 
\begin{eqnarray}
G_{s}<\bar{q}\lambda ^{a}q>=g_{s}\sigma ^{a} &  & \\
G_{v}<\bar{q}\lambda ^{a}\gamma _{\mu }q>=g_{v}V_{\mu }^{a} &  & 
\end{eqnarray}
The similar pattern may be extended to the axial mesons 
\begin{eqnarray}
G_{s}<\bar{q}\lambda ^{a}i\gamma _{5}q>=g_{s}\phi ^{a} &  & \\
G_{v}<\bar{q}\lambda ^{a}\gamma _{\mu }i\gamma _{5}q>=g_{v}A_{\mu }^{a} &  & 
\end{eqnarray}
Mesons masses are defined as 
\begin{eqnarray}
m_{\sigma}=g_{s}/\sqrt{G_{s}} &  & \\
m_{v}=g_{v}/2\sqrt{G_{v}} &  & 
\end{eqnarray}
 This is a process of bosonization in which the NJL model produce essentially
the \( u(3) \) linear sigma model as an approximate effective theory
for the scalar and pseudoscalar meson sector \cite{torn}.

\begin{figure}
{\centering \resizebox*{15cm}{!}{\includegraphics{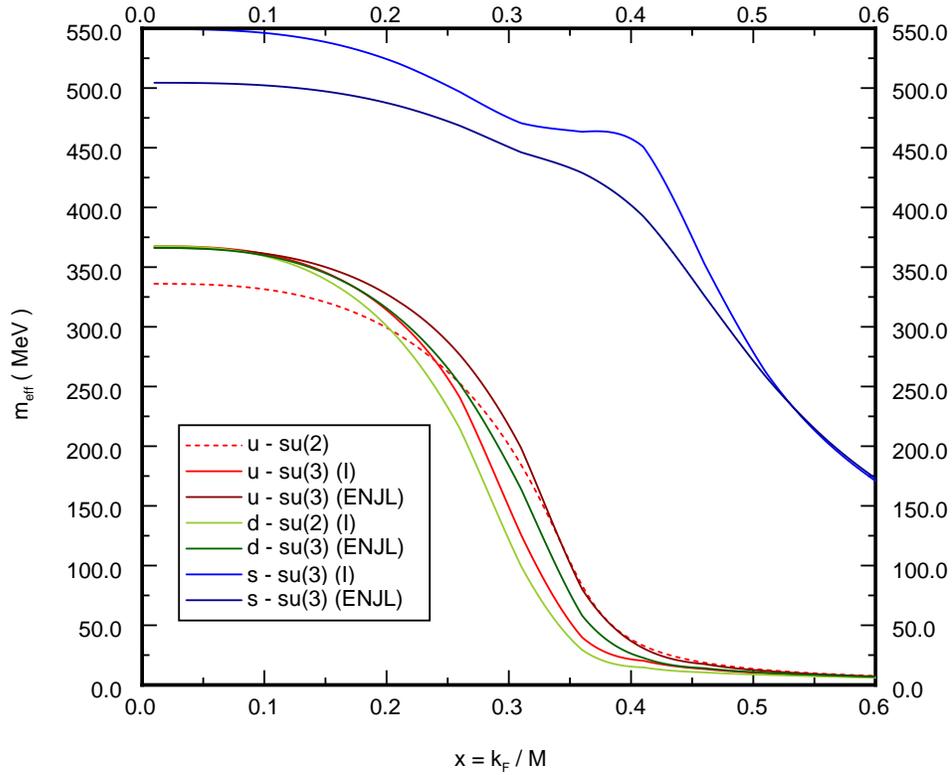}} \par}

\caption{The effective quark masses for different NJL models as a function
of the \protect\( u\protect \) quark dimensionless Fermi momentum
\protect\( x=k_{F,u}/M\protect \) (\protect\( M=939\, MeV\protect \)
is the nucleon mass)}

\label{figd}
\end{figure}

\begin{figure}
{\centering \resizebox*{15cm}{15cm}{\includegraphics{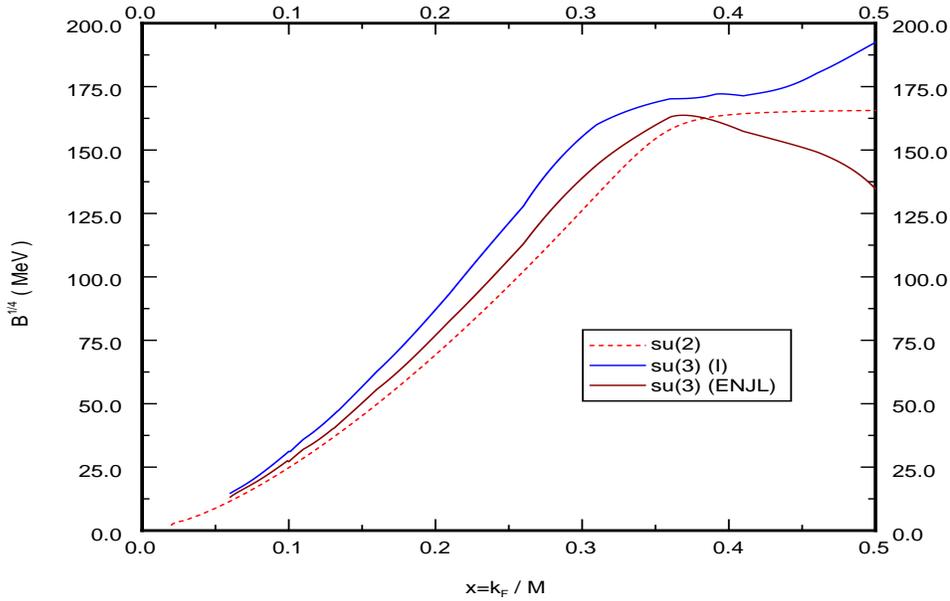}} \par}
\caption{\label{figb}The effective bag constant as a function of the \protect\( u\protect \)
quark dimensionless Fermi momentum \protect\( x=k_{F,u}/M\protect \)
(\protect\( M=939\, MeV\protect \) is the nucleon mass).}
\end{figure}

In this paper the variational method based on the Feynman-Bogolubov
inequality \cite{rm} is incorporated (see more details in \cite{rm1})
\begin{equation}
\Omega \leq \Omega _{1}=\Omega _{0}(m_{eff})+<H-H_{0}>_{0}.
\end{equation}
 with   the trial Lagrange function described by
\begin{equation}
{\mathcal{L}}_{0}(m_{eff})=\overline{q}(i\gamma ^{\mu }\bar{D}_{\mu }-m_{eff})q
\end{equation}
and suggested by the mean field form of the Lagrange function (\ref{lagsug}).
The covariant derivative
\begin{equation}
\bar{D}_{\mu }=\partial _{\mu }+\frac{1}{2}ig_{v}V_{\mu }^{a}\lambda ^{a}
\end{equation}
is limited to the commuting Cartan subalgebra \( \lambda ^{i}=\{\, \lambda ^{0},\, \lambda ^{3},\, \lambda ^{8}\, \} \).
This approach introduces the fermion interaction with homogeneous boson condensates
\( \sigma ^{a},\, V^{a}_{\mu } \) which together with the effective
masses \( m_{eff} \) will be treated as variational parameters. \( \Omega _{0} \)
is the thermodynamic potential of the effectively free quasiparticle
system 
\begin{eqnarray}
\Omega _{0}(m_{eff})=E_{0}-k_{B}T\frac{N_{q}}{2\pi ^{2}}\sum _{f}\int _{0}^{\Lambda }dkk^{2}(ln(1+e^{-\beta (\sqrt{k^{2}+m_{eff,f}^{2}}-\mu _{f})})+ &  & \nonumber \\
ln(1+e^{-\beta (\sqrt{k^{2}+m_{eff,f}^{2}}+\mu _{f})})) &  & 
\end{eqnarray}
We could not forget about energy of quantum fluctuations because it
depends on the quark effective mass. As fermions give \( -1/2\, \hbar \omega  \)
to the vacuum energy, we get 
\begin{equation}
\mathcal{E}_{0}=-\, \frac{\, N_{q}}{2\pi ^{2}}\sum _{f=\{u,d,s\}}\int _{0}^{\Lambda }dkk^{2}\sqrt{k^{2}+m_{eff,f}^{2}}
\end{equation}
assuming that if \( m_{eff}=m_{0} \) the energy of quantum fluctuations
may be neglected. The effective quark masses entering into the Lagrangian
function \( L_{0}(m_{eff}) \) of the trial system are calculated
from the extremum conditions 
\begin{equation}
\label{efef}
\frac{\partial \Omega _{1}}{\partial m_{eff,f}}=0
\end{equation}
which give\begin{eqnarray*}
 & (m_{eff})_{f,f'}=m_{eff,f}\delta _{f,f'}=m_{c}\delta _{f,f'}- & \\
 & G_{s}\sum _{a=0}^{8}<\bar{q}_{f}\lambda ^{a}q_{f}>_{0}\lambda ^{a}_{f,f'}+2K\delta _{f,f'}\prod _{f'\neq f}<\bar{q}_{f'}q_{f'}>_{0} & 
\end{eqnarray*}
or
\begin{equation}
\label{m1}
m_{eff,f}=m_{c,f}-\bar{G}_{s}<\bar{q}_{f}q_{f}>_{0}+2K\prod _{f'\neq f}<\bar{q}_{f'}q_{f'}>_{0},
\end{equation}
where
\begin{eqnarray}
 & <\overline{q}_{f}q_{f}>_{0}=\frac{m_{eff,q}\, N_{q}}{\pi ^{2}}\int _{0}^{\Lambda }\frac{k^{2}dk}{\sqrt{k^{2}+m_{eff,f}^{2}}}\{\frac{1}{\exp (\beta (\sqrt{k^{2}+m_{eff,f}^{2}}-\mu _{f}))+1}+ & \nonumber \\
 & \frac{1}{\exp (\beta (\sqrt{k^{2}+m_{eff,f}^{2}}+\mu _{f}))+1}-1\}. & 
\end{eqnarray}
At \( T=0 \) we have only 
\begin{equation}
<\bar{q}_{f}q_{f}>_{0}=-m_{eff,f}\, \frac{N_{q}}{2\pi ^{2}}\int _{k_{F}}^{\Lambda }dk\frac{k^{2}}{\sqrt{k^{2}+m_{eff,f}^{2}}}.
\end{equation}
In vacuum we get the constituent quarks with mass
\begin{equation}
m_{c}=m_{0}-2\bar{G}_{s}<\bar{q}q>_{0v}+2K\prod _{f'=\{u,d,s\}}<\bar{q}_{f'}q_{f'}>_{0v}
\end{equation}
where
\begin{equation}
<\bar{q}_{f}q_{f}>_{0v}=-m_{v,f}\frac{N_{q}}{2\pi ^{2}}\int _{0}^{\Lambda }dk\frac{k^{2}}{\sqrt{k^{2}+m_{v,f}^{2}}}
\end{equation}
 At minimum the effective free energy has the form 
 \begin{equation}
\Omega _{eff}=\Omega _{1}|_{min}=\Omega _{0}(m_{eff})+B_{eff}
\end{equation}
with the effective bag constant 
\begin{eqnarray}
B_{eff}=\frac{1}{2}G_{s}<\bar{q}\lambda ^{a}q>^{2}-4K\prod _{f=\{u,d,s\}}<\bar{q}_{f}q_{f}> &  & \nonumber \\
-\frac{1}{2}G_{v}<\bar{q}\gamma ^{\mu }\lambda ^{a}q><\bar{q}\gamma _{\mu }\lambda ^{a}q>-B_{0} &  & \label{beffe} 
\end{eqnarray}
Quarks as effectively free quasiparticles in vacuum with nonvanishing
bag 'constant'. The constant \( B_{0} \) was chosen in this way to
have massive (\( m_{v,u,d} \) \( = \) \( 367.61 \) \( MeV \),
\( m_{v,s} \) \( = \) \( 549.45 \) \( MeV \) for the NJL (I) parameters
set and \( m_{v,u,d}= \) \( 366.13 \) \( MeV \), \( m_{v,s} \)
\( = \) \( 504.13 \) \( MeV \) for enlarged NJL model). However,
in  high density medium  they are less massive (Figure \ref{figd})
but the effective bag constant (Figure \ref{figb})  grows up
to \( B_{eff}^{1/4}\div 150-180\, MeV \). The frequently used case
with current quarks and bag constant is valid only in very high
density limit. This is a justification when the quark matter phase
is modeling in the context of the MIT bag model \cite{GlenBook,FarhiJaffe84,MITbag}
as a Fermi gas of \( u \), \( d \), and \( s \) quarks. In this
model the phenomenological bag constant \( B_{\textrm{MIT}} \) is
introduced to mimic QCD interactions. In the original MIT bag model
the bag constant was constant and the value \( B=B_{c}=(154.5\, MeV)^{4} \)
makes the strange matter absolutely stable. 

To avoid quantum fluctuations the meson fields may be redefined to
produce the phenomenological sigma field as
\begin{equation}
g_{s}\, \varphi _{a}=G_{s}(<\bar{q}\lambda ^{a}q>_{0}-<\bar{q}\lambda ^{a}q>_{0v})
\end{equation}
so, the effective quark mass can be rewritten as
\begin{equation}
m_{eff}=m_{c}-g_{s}\, \varphi _{a}\lambda ^{a}
\end{equation}
Thus, the effective quark mean field (QMF) theory appears. The Lagrange
function in the mean field approximation may be written in the following
form 
\begin{eqnarray}
{\mathcal{L}}_{QMF}=\overline{q}(i\gamma ^{\mu }D_{\mu }
-m_{c})q+\frac{1}{2}\partial _{\mu }\varphi _{a}\partial ^{\mu }\varphi _{a}-\frac{1}{2}m_{\sigma}^{2}\varphi _{a}\varphi _{a} &  & \\
-\frac{1}{4}F_{\mu \nu }^{a}F^{a,\mu \nu }+\frac{1}{2}m_{v}^{2}V_{\mu }^{a}V^{a\mu } &  & \label{qlef} 
\end{eqnarray}
Unfortunately, the vacuum quantum fluctuations are missed, but meson
fields gain the dynamical character. Restricted ourself only to \( u(2)\times u(1) \)
subalgebra (\( a=\{0,1,2,3,8\} \)) case we have the simplest version
of the QMF theory. Defining new base with \( \tau ^{a},\, \, a=\{0,1,2,3,4\} \)
as 
\begin{equation}
\tau ^{0}=\left( \begin{array}{ccc}
1 & 0 & 0\\
0 & 1 & 0\\
0 & 0 & 0
\end{array}\right) \, \, \tau ^{i}=\left( \begin{array}{cc}
\sigma ^{i} & 0\\
0 & 0
\end{array}\right) \, \, for\, \, \, \tau ^{4}=\sqrt{2}\left( \begin{array}{ccc}
0 & 0 & 0\\
0 & 0 & 0\\
0 & 0 & 1
\end{array}\right) 
\end{equation}
$(i=\{1,2,3\})$ the meson fields may be decomposed as follows
\[
\varphi =\varphi _{a}\tau ^{a}=\sigma \tau ^{0}+\delta _{i}\tau ^{i}+\sigma _{s}\tau ^{4}
\]
and 
\[
V_{\mu }=V^{a}_{\mu }\lambda ^{a}=\omega _{\mu }\tau ^{0}+\omega _{s,\mu }\tau ^{4}+b_{i}\tau ^{i}
\]
Now the new meson fields are denoted by 
\[
\omega _{\mu }=\sqrt{\frac{2}{3}}V^{0}_{\mu }+\frac{1}{\sqrt{2}}V^{8}_{\mu },
\]
\[
\omega _{s,\mu }=\sqrt{\frac{1}{3}}V^{0}_{\mu }-V^{8}_{\mu },
\]
and \( b_{\mu }^{i}=V_{\mu }^{i} \) (with \( i=\{1,2,3\} \)), respectively.
The simplest u(2) version (\( \sigma _{s}=0 \), \( \omega _{s,\mu }=0 \))
has the Lagrange density function with the following form 
\begin{eqnarray}
{\mathcal{L}}_{QMF} 
 & =\frac{1}{2}\partial _{\mu }\sigma \partial ^{\mu }\sigma -U(\sigma )-\frac{1}{4}\Omega _{\mu \nu }\Omega ^{\mu \nu }+\frac{1}{2}M_{\omega }^{2}\omega _{\mu }\omega ^{\mu }+ & \nonumber  \\
 & \frac{1}{2}\partial _{\mu }\delta _{i}\partial ^{\mu }\delta _{i}-\frac{1}{2}m_{\delta }^{2}\delta _{i}^{2}-\frac{1}{4}R_{\mu \nu }^{a}R^{a\mu \nu }+\frac{1}{2}M_{\rho }^{2}b^{a}_{\mu }b^{a\mu }+ & \nonumber \\
 & \overline{q}(i\gamma ^{\mu }D_{\mu }-m_{c})q+g^{q}_{\sigma }\sigma \overline{q}q+g^{q}_{\delta }\delta _{i}\overline{q}\tau ^{i}q 
 \label{lag}
\end{eqnarray}
The field tensors \( R_{\mu \nu }^{a} \), \( \Omega _{\mu \nu } \)
and the covariant derivative \( D_{\mu } \) are given by 
\begin{equation}
R_{\mu \nu }^{a}=\partial _{\mu }b^{a}_{\nu }-\partial _{\nu }b^{a}_{\mu }+g_{\rho }\varepsilon _{abc}b_{\mu }^{b}b_{\nu }^{c},
\end{equation}
\begin{equation}
\Omega _{\mu \nu }=\partial _{\mu }\omega _{\nu }-\partial _{\nu }\omega _{\mu },
\end{equation}
\begin{equation}
D_{\mu }=\partial _{\mu }+\frac{1}{2}ig_{\rho }b^{a}_{\mu }\tau ^{a}+\frac{1}{2}ig_{\omega }\omega _{\mu }.
\end{equation}
 The \( \delta =0 \) limit gives the simplest version of the Quark
Mean Field (QMF) model. 

Some years ago Guichon proposed an interesting model concerning  the change
of the nucleon properties in nuclear matter (quark-meson coupling
model (QMC)) \cite{aust}. The model construction mimics the relativistic
mean field theory, where the scalar \( \sigma  \) and the vector
meson \( \omega  \) fields couple not with nucleons but directly
with quarks. The quark mass has to change from its bare current mass
due to the coupling to the \textbf{\( \sigma  \)} meson. More recently,
Shen and Toki \cite{toki1} have proposed a new version of the QMC
model, where the interaction takes place between constituent quarks
and mesons. They refer the model as the quark mean field model (QMF).
In this work we shall also investigate the quark matter within the
QMF theory motivated by parameters coming from the enlarged Nambu-Jona-Lasinio
(ENJL). Enlargement of the NJL model is based on inclusion of vector
mesons while QMF model includes vector mesons at the beginning. 

Here the QMF model is a bit generalized by the inclusion of  the isovector \( \delta  \)
\( (a_{0}(980)) \) meson. It splits \( u \) and \( d \) masses
(or proton and neutron masses in the case of RMF approach \cite{kub4,liu3}).
Both \( \delta ^{i} \) and \( b_{\mu }^{i} \) mesons may be neglected
in the case of symmetric nuclear matter. Its role in the asymmetric nuclear
matter of the neutron star is significant and is a subject of our current
interest. 

The QMF model is more flexible. The \( SU(3) \) symmetry restrict
\( g^{q}_{\sigma }=\sqrt{2/3}g_{s} \), \( g^{q}_{\omega }=g_{\omega } \),
\( g^{q}_{\rho }=g_{\rho } \) to \( g_{\rho }=g_{\omega }=g_{v} \)
and \( m_{\rho }=m_{\omega }=m_{v} \).

\begin{figure}
{\centering \resizebox*{15cm}{!}{\includegraphics{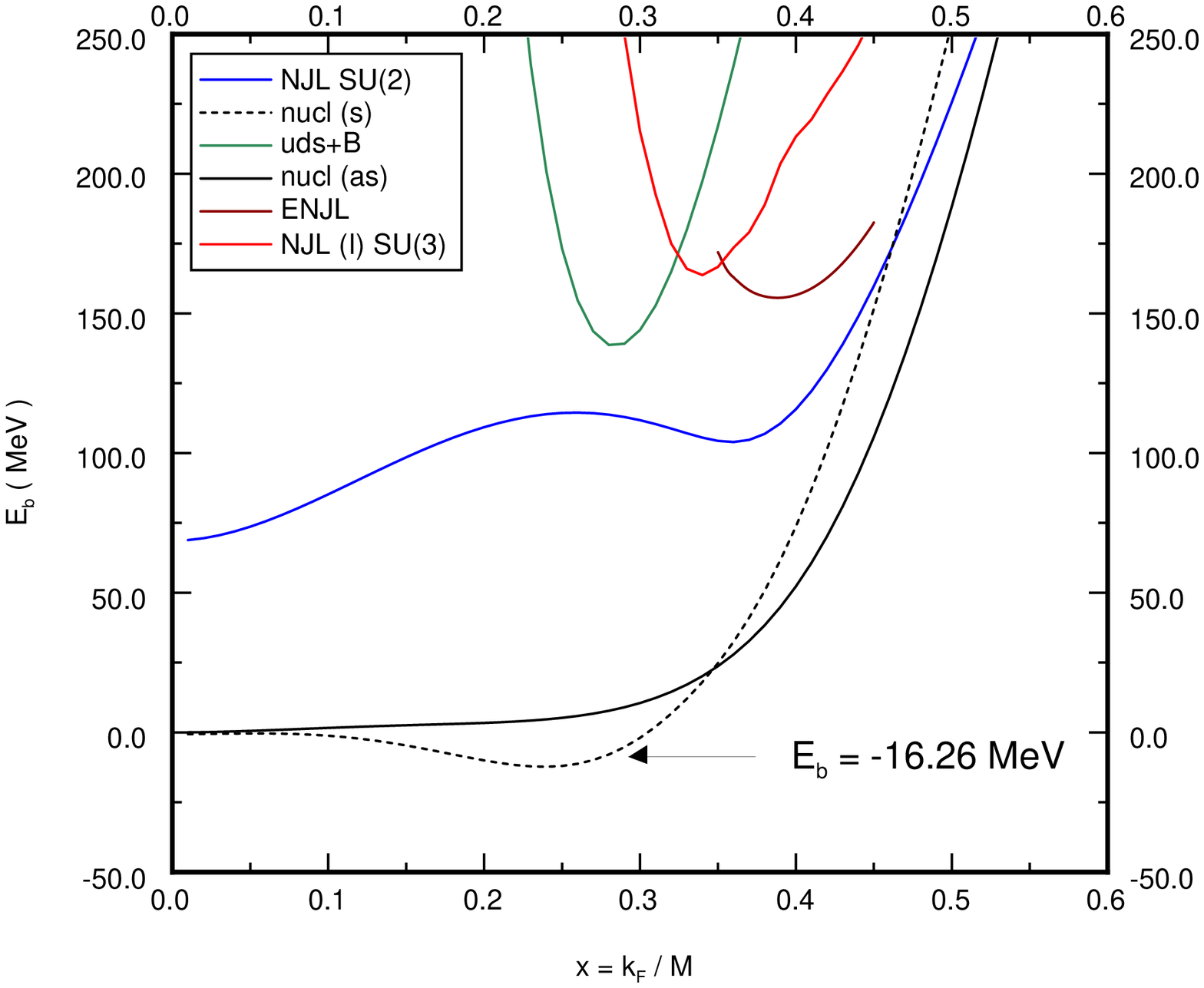}} \par}
\caption{\label{figbin}The binding energy \protect\( E_{b}\protect \) for
the quark and nuclear system. The pure quark system (\protect\( usd\protect \)
quarks with the bag constant \protect\( B\protect \)) and QMF have
nonrealistic behavior for small densities as \protect\( B\neq 0\protect \).
To compare the symmetric nuclear matter (the RMF approach with TM1\cite{tm1}
parameterization) is presented. }
\end{figure}
The effective quarks mass \( m_{F,f} \) (or \( \delta _{f}=m_{F,f}/M \)
) dependence on the dimensionless Fermi momentum \( x=k_{F,u}/M \)
is presented on the Figure \ref{figd}. 
\begin{figure}
{\centering \resizebox*{15cm}{15cm}{\includegraphics{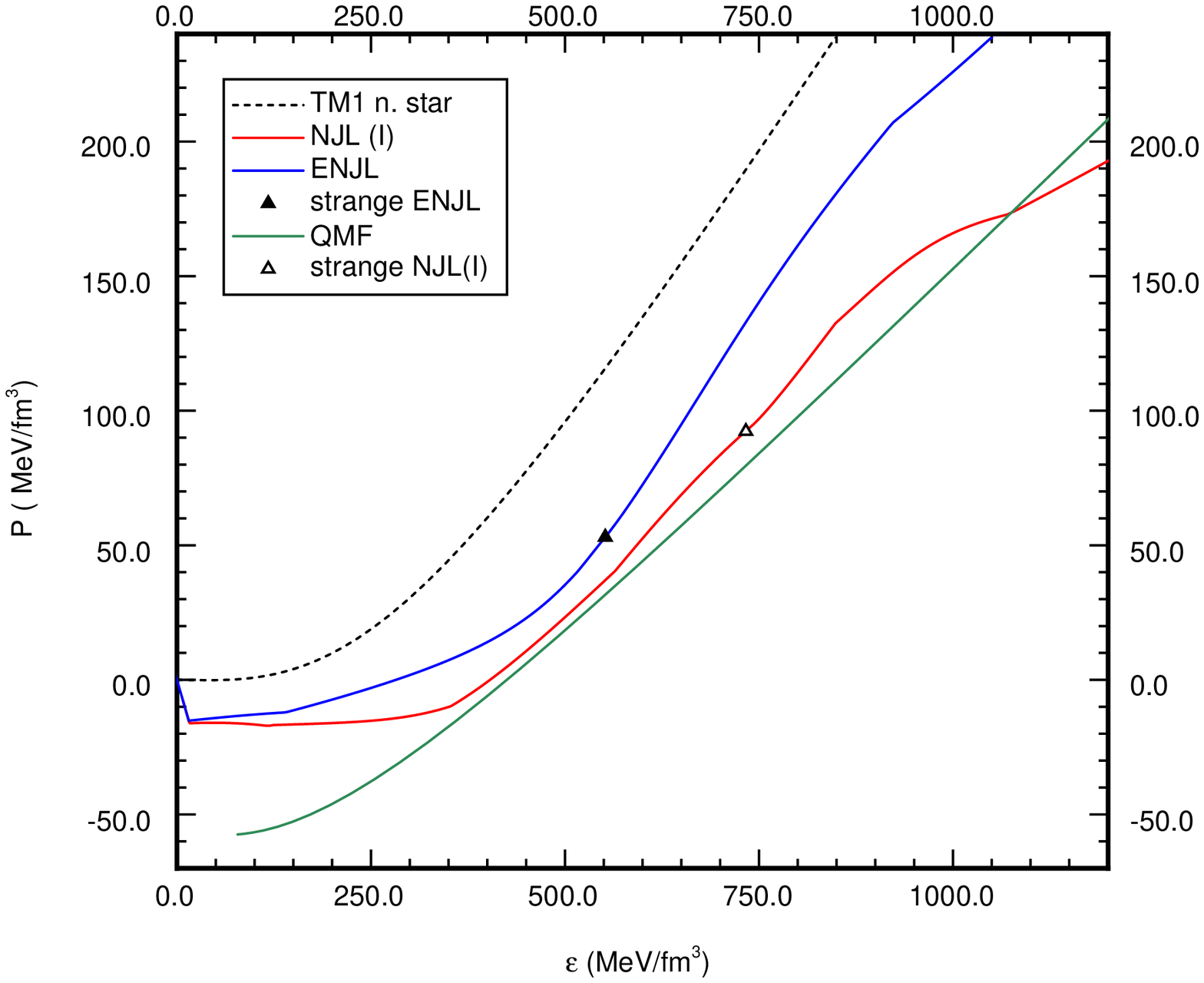}} \par}
\caption{\label{fesob}The form of equation of state (EoS) for quark ( NJL(I),
enlarged NJL and QMF model) matter and nucleon one (TM1\cite{tm1}).
The blue (\protect\( n^{s}_{B}=3.94\, n_{B}^{0}\protect \)) points
the strange \protect\( s\protect \) quark appearence in the strange
star. }
\end{figure}
\begin{figure}
{\centering \resizebox*{15cm}{15cm}{\includegraphics{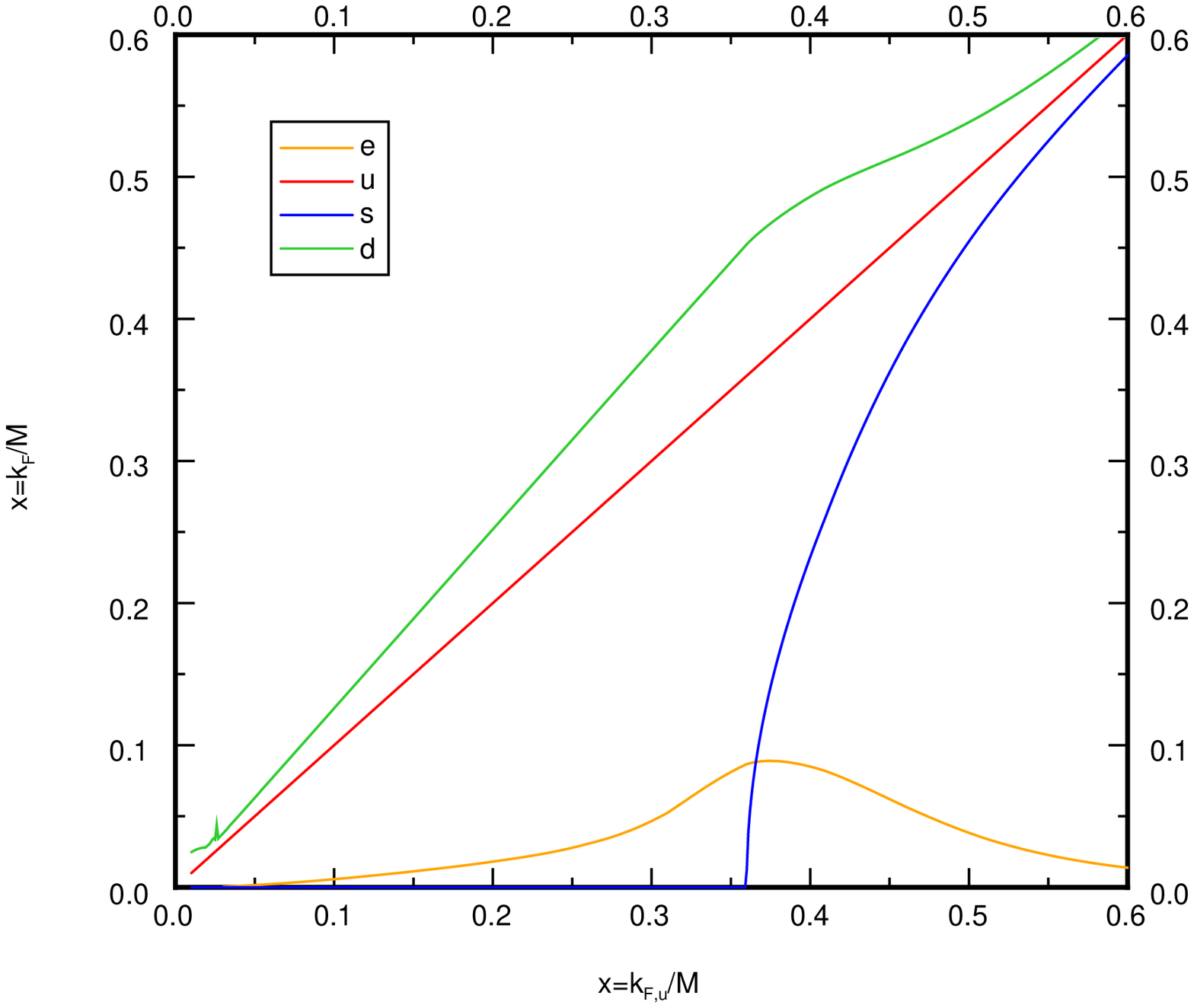}} \par}
\caption{\label{figxs}The electron and quarks dimensionless Fermi momentum
as a function of the \protect\( u\protect \) quark one (\protect\( x=k_{F,u}/M\protect \),
\protect\( M=939\, MeV\protect \) is the nucleon mass). The muon
distribution is not visible in this scale.}
\end{figure}

There is not quarks confinement both in the NJL and QMF models. There
is no mechanism (exept for the NJL solvable model \cite{blaschke})
to prevent hadrons to decay into free constituent quarks. Free constituent
quark will produce nearly the same density and pressure like free
nucleons \( 3m_{v,u,d}\sim M \). Without any mechanism of confinement
the quark star for small densities will have properties very similar
to those  of neutron stars (the case \( x_{v}>0.65 \) in paper
\cite{Hana01}) or even white dwarfs. However, this is rather unphysical
artefact. It is natural to assume that quarks are not allowed to propagate
over the distance \( \lambda \sim m_{eff}^{-1}. \) In this language
the confinement mechanism introduces infrared cut-off \( \lambda  \)
\cite{blaschke1}. The quarks confinement mechanism in the form of
the harmonic oscillator potential \cite{toki1} may give the nucleon
mass \( M=M(\sigma )=M-g_{N\sigma }\sigma +... \) and generate the
Relativistic Mean Field (RMF) approach. 

Relativistic Mean Field Theory \cite{wal} is very useful in describing
nuclear matter and finite nuclei. Recent theoretical studies show
that the properties of nuclear matter can be described nicely in terms
of the Relativistic Mean Field Theory. Properties of the neutron star
in this model were also examined \cite{Weber,GlenBook,toki,rm1,rm2}. 

Its extrapolation to large charge asymmetry is of considerable interest
in nuclear astrophysics and particulary in constructing neutron star
model where extreme conditions of isospin is realized. The construction
of neutron star model is based on various realistic equations of state
and results in the general picture of neutron stars interiors.Thus
the proper form of the equation of state is essential in determining
neutron star properties such as the mass-range, mass-radius relation
or the crust thickness. However, the complete and more realistic description
of a neutron star requires taking into consideration not only the
interior region of a neutron star but also the remaining layers, namely
the inner and outer crust and the surface. ~The Lagrangian of the
RMF theory which tends towards the construction of neutron star model
contains baryon and meson degrees of freedom and as input quantities
coupling constants of the mesons and parameters of the potential \( U(\sigma ) \)
which are determined from nuclear matter properties. 
\begin{eqnarray}
{\mathcal{L}}_{QMF} & =\frac{1}{2}\partial _{\mu }\sigma \partial ^{\mu }\sigma -U(\sigma )-\frac{1}{4}\Omega _{\mu \nu }\Omega ^{\mu \nu }+\frac{1}{2}M_{\omega }^{2}\omega _{\mu }\omega ^{\mu }+ & \nonumber  \\
 & -\frac{1}{4}R_{\mu \nu }^{a}R^{a\mu \nu }+\frac{1}{2}M_{\rho }^{2}b^{a}_{\mu }b^{a\mu }+\frac{1}{4}c_{3}(\omega _{\mu }\omega ^{\mu })^{2} & \nonumber \\
 & \overline{\psi }(i\gamma ^{\mu }D_{\mu }-M)\psi +g_{N\sigma }\sigma \overline{\psi }\psi 
\end{eqnarray}
Now \( \psi  \) describes nucleons and \( g_{N\sigma }=3g^{q}_{s}, \)
\( g_{N\omega }=3g^{q}_{\omega } \) and \( g_{N\rho }=g^{q}_{\rho }, \). 

The potential function \( U(\sigma ) \) may possesses a polynomial
form introduced by Boguta and Bodmer \cite{Bogut77} in order to get
a correct value of the compressibility \( K \) of nuclear matter
at saturation density 
\begin{equation}
U(\sigma )=\frac{1}{2}m^{2}_{\sigma}\sigma ^{2}+\frac{1}{3}g_{2}\sigma ^{3}+\frac{1}{4}g_{3}\sigma ^{4}.
\end{equation}

Different parameter sets give different forms of the equation of state
in the high density region above saturation. In this paper the \( TM1 \)
\cite{tm1} parameter set was exploit. The RMF model leads its own
fenomenological life but its parameters should be connected to the
enlaged NJL model. The \( TM1 \) parametrization suggests that \( x_{v}=0.65 \). 

The \( \rho  \) meson plays decisive role in accounting for the asymmetry
energy of nuclear matter thus its inclusion in a theory of neutron
star matter is essential. Also the proton number density is determined
by this meson. The results for the binding enrgy is presented
in Figure \ref{figbin}. It shows that the nuclean or quark matter
in \( \beta  \)-equilibrium has a larger energy per particle than
symmetric nuclear matter. For neither parameter set are these matters
self-bounded. Figure \ref{figbin} depicts the binding energy for
different models. It reproduces the standard results for symmetric
nuclear matter too. The asymmetric matter is less bound than the symmetric
one. In this model we are dealing with the electrically neutral neutron
or quark star matter being in \( \beta  \)-equilibrium. Therefore
the imposed constrains, namely the charge neutrality and \( \beta  \)-equilibrium,
imply the presence of leptons.

\section*{\textcolor{blue}{The quark star properties}}

To calculate the properties of the quark star we need the energy-momentum
tensor. The energy-momentum tensor can be calculated taking the quantum
statistical average 
\begin{eqnarray}
\bar{T}_{\mu \nu }=<T_{\mu \nu }>,
\end{eqnarray}
where 
\begin{eqnarray}
T_{\mu \nu }=2\frac{\partial {{\mathcal{L}}}}{\partial g^{\mu \nu }}-g_{\mu \nu }{\mathcal{L}}. & 
\end{eqnarray}
 In case of the fermion fields it is more convenient to use the reper
field \( e^{a}_{\mu } \) defined as follows \( g_{\mu \nu }=e^{a}_{\mu }e^{b}_{\nu }\eta _{ab} \)
where \( \eta _{ab} \) is the flat Minkowski space-time matrix. Then
\begin{equation}
T_{\mu \nu }=e_{\mu }^{a}\frac{\partial {\mathcal{L}}}{\partial e^{a\nu }}-g_{\mu \nu }{\mathcal{L}}.
\end{equation}
We define the density of energy and pressure by the energy - momentum
tensor 
\begin{equation}
\label{tep}
\bar{T}_{\mu \nu }=(P+\epsilon )u_{\mu }u_{\nu }-Pg_{\mu \nu }=\left( \begin{array}{cccc}
\epsilon =c^{2}\rho  & 0 & 0 & 0\\
0 & P & 0 & 0\\
0 & 0 & P & 0\\
0 & 0 & 0 & P
\end{array}\right) 
\end{equation}
where \( u_{\mu } \) is a unite vector (\( u_{\mu }u^{\mu }=1 \)).
Both \( \epsilon  \) and \( P \) depend on the quarks chemical potential
\( \mu  \) or Fermi momentum \( x_{f} \). Fermions (quarks and leptons)
contribution to the energy and pressure are
 \begin{equation}
\epsilon _{F}=\sum _{f=\{u,d,s\}}\epsilon _{0}\chi _{B}(x_{f},T)+\sum _{f=\{e,\mu \}}\epsilon _{0}\chi _{L}(x_{f},T)
\end{equation}
\begin{equation}
P_{F}=\sum _{f=\{u,d,s\}}P_{0}\phi (x_{f},T)+\sum _{f=\{e,\mu \}}P_{0}\phi (x_{f},T)
\end{equation}
The fact that effective quarks mass depends \( m_{eff,f}=\delta _{f}M \)
on fermion concentration ( or quark chemical potential \( \mu _{f} \))
now must be included into \( \chi (x_{f},T) \) and \( \phi (x_{f},T) \),
\begin{eqnarray}
\chi (x,T)=\frac{3}{^{\pi ^{2}}}\int _{\lambda }^{\Lambda /M}dz\, z^{2}\sqrt{z^{2}+\delta ^{2}(x)}\{\frac{1}{\exp ((\sqrt{\delta ^{2}(x)+z^{2}}-\mu ')/\tau )+1} &  & \\
+\frac{1}{\exp ((\sqrt{\delta ^{2}(x)+z^{2}}+\mu ')/\tau )+1}, &  & \nonumber 
\end{eqnarray}
\begin{eqnarray}
\phi (x,T)=\frac{1}{\pi ^{2}}\int _{\lambda }^{\Lambda /M}\frac{z^{4}dz}{\sqrt{z^{2}+\delta ^{2}(x)}}\{\frac{1}{\exp ((\sqrt{\delta ^{2}(x)+z^{2}}-\mu ')/\tau )+1} &  & \\
+\frac{1}{\exp ((\sqrt{\delta ^{2}(x)+z^{2}}+\mu ')/\tau )+1}\} &  & \nonumber 
\end{eqnarray}
 where \( \tau =(k_{B}T)/M \), 
\begin{equation}
\label{mu}
\mu '=\mu /M=\sqrt{\delta ^{2}(x)+x^{2}}
\end{equation}
 and 
\begin{equation}
\label{mui}
x=k/M
\end{equation}
for each flaver \( f \). Similar to paper \cite{toki} we have introduced
(\ref{mu},\ref{mui}) the dimensionless {}``Fermi'' momentum even
at finite temperature which exactly corresponds to the Fermi momentum
at zero temperature. To avoid free quark contributions to the equation
of state coming from small densities the infrared cut-off \( \lambda =\delta  \)
\cite{blaschke1} was introduced. The case \( \lambda =0 \) with
the NJL (I) parameters set nicely reproduce the result od paper \cite{Hana01}. 

The parametric dependence
on \( \mu  \) (or \( x_{f} \)) defines the equation of state. The
various equations of state for different parameters sets are presented
in Figure \ref{fesob}. The binding energies 
\[
E_{b}=\rho /n_{B}-Mc^{2}
\]
for the bulk nuclear (\( n_{B} \) is the baryon number density, \( n_{B}=(n_{p}+n_{n}) \))
and quark matter (\( n_{B}=(n_{u}+n_{d}+n_{s})/3 \)) are presented
in  Figure \ref{figbin}.

The metric is static, spherically symmetric and asymptotically flat
\begin{equation}
\label{tensor metryczny}
g_{\mu \nu }=\left( \begin{array}{cccc}
e^{\nu (r)} & 0 & 0 & 0\\
0 & -e^{\lambda (r)} & 0 & 0\\
0 & 0 & -r^{2} & 0\\
0 & 0 & 0 & -r^{2}sin^{2}\theta 
\end{array}\right) 
\end{equation}
 (where \( \nu (r) \) and \( \lambda (r) \) are functions of a radius
\( r \) ). Einstein equations (in isotropic case) leads to the standard
Tolman-Oppenheimer-Volkoff equations \cite{tov}. The equations describing
masses and radii of quark star are determined by the proper form of
the equation of state. The obtained form of the equation of state
is the base for calculating macroscopic properties of the star. In
order to construct the mass-radius relation for given form of the
equation of state the OTV equations have to be solved 
\begin{equation}
\label{teq1}
\frac{dP(r)}{dr}=-\frac{G}{r^{2}}(\rho (r)+\frac{P(r)}{c^{2}})\frac{(m(r)+\frac{4\pi }{c^{2}}P(r)r^{3})}{(1-\frac{2Gm(r)}{c^{2}r})}
\end{equation}
\begin{equation}
\label{teq2}
\frac{dm(r)}{dr}=4\pi r^{2}\rho (r)
\end{equation}
 The continuity condition for the energy-momentum tensor \( T_{;\nu }^{\mu \nu }=0 \)
defines the connection between gravitational potential \( \nu (r) \)
(\ref{tensor metryczny}) and pressure and density profiles \( P(r) \)
and \( \rho (r) \) 
\begin{equation}
\label{teq}
\frac{d\nu (r)}{dr}=-\frac{2}{P(r)+c^{2}\rho (r)}\frac{dP(r)}{dr}.
\end{equation}
The equation (\ref{teq2}) determines the  function \( \lambda (r) \)
\begin{eqnarray*}
e^{-\lambda (r)}=1-\frac{2Gm(r)}{r}.\label{leq} 
\end{eqnarray*}
Having solved the OTV equation the pressure \( P(r) \), mass \( m(r) \)
and density \( \rho (r) \) profile is obtained. To obtain the total
radius \( R \) of the star the fulfillment of the condition \( P(R)=0 \)
is necessary. Introducing the dimensionless variable \( \xi , \)
which is connected with the star radius \( r \) by the relation \( r=a\xi  \)
enables to define the functions \( P(r) \), \( \rho (r) \) and \( m(r) \)
\begin{equation}
\rho (r)=\rho _{0}\chi (x(\xi ))
\end{equation}
\begin{equation}
P(r)=P_{0}\varphi (x(\xi ))
\end{equation}
\begin{equation}
m(r)=M_{\odot }u(\xi )
\end{equation}
 by \( \xi  \). Dimensionless functions defined as 
\begin{equation}
\alpha _{0}=\frac{GM_{\odot }\rho _{c}}{P_{0}a},\, \, \beta _{0}=3\frac{M_{s}}{M_{\odot }},\, \, M_{s}=\frac{4}{3}\pi \rho _{0}a^{3}
\end{equation}
 are needed to achieve the OTV equation of the following form 
\begin{equation}
\label{volk1}
\frac{d\varphi }{d\xi }=-\alpha _{0}(\chi (x(\xi ))+\varphi (x(\xi )))\frac{u(\xi )+\beta _{0}\varphi (x(\xi ))\xi ^{3}}{\xi ^{2}(1-\frac{r_{g}}{a}\frac{u(\xi )}{\xi })}
\end{equation}
\begin{equation}
\label{volk2}
\frac{du}{d\xi }=\beta _{0}\chi (x(\xi ))\xi ^{2}
\end{equation}
 with \( r_{g} \) being the gravitational radius. The equations (\ref{volk1},\ref{volk2})
are easy integrated numerically. These are equations for dimensionless
mass \( u(\xi )=m(r)/M_{\odot } \) up to radius dimensionless \( \xi  \)
and the quark \( u \) dimensionless Fermi momentum \( x=k_{F,u}/M \).
Knowing the variable \( x \) all star properties can be calculated. Quarks
and electron dimensionless Fermi momenta dependences  on \( x \) is presented in Figure \ref{figxs}. 

Both nuclear and quark matter being in \( \beta  \)-equilibrium are
not bound (Figure \ref{figbin}). For quark matter at moderate densities is bound due to the presence  
the bag constant \( B \) which acts as  a negative pressure \( P=-B+... \)
(Figure \ref{fesob}).Higher density matter is bound  only
gravity. The gravitational binding energy of the star is defined
as 
\begin{equation}
E_{b,g}=(M_{p}-m(R))c^{2}
\end{equation}
where 
\begin{equation}
M_{p}=4\pi \int _{0}^{R}drr^{2}(1-\frac{2Gm(r)}{c^{2}r})^{-\frac{1}{2}}\rho (r)
\end{equation}
is the proper star mass.

\begin{figure}
{\centering \resizebox*{15cm}{15cm}{\includegraphics{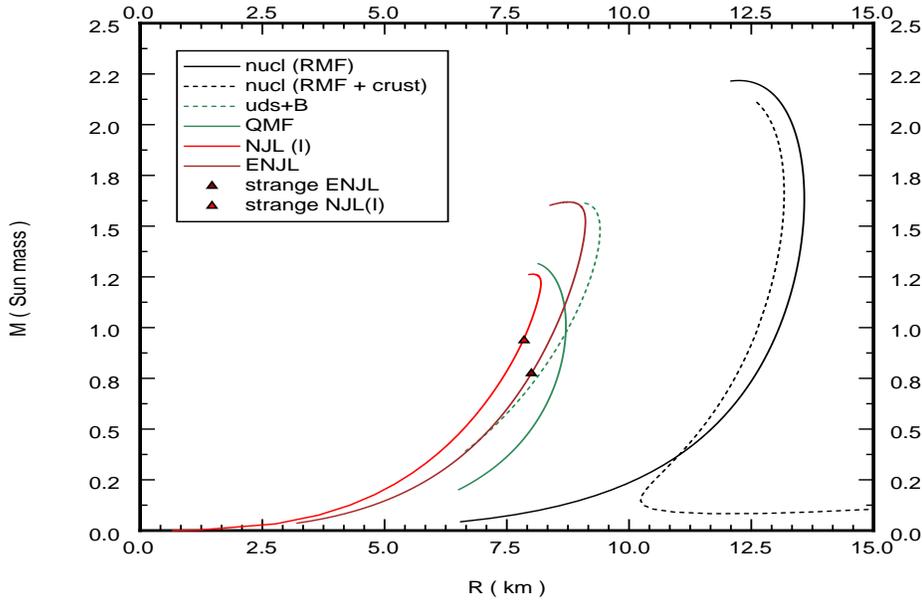}} \par}
\caption{\label{figrm}The mass - radius \protect\( M(R)\protect \) dependence
for the quark star(NJL(I), ENJL (\protect\( x_{v}=1.06\protect \),
the solid brown line and \protect\( x_{v}=0.65\protect \), the solid
violet curve), QMF model and pure \protect\( uds\protect \) matter
(QMC)). To compare this relation to the neutron star in the RMF approach
it is presented as the black line (TM1 - solid line) for pure \protect\( npe\protect \)
nuclear matter and (TM1+Bonn+Negele+Vautherin) - dotted line) with
crust.}
\end{figure}

Strange quarks star in the NJL model are rather small in comparison
to the neutron ones. To compare these strange stars to  neutron
stars models obtained in  the RMF approach the mass - radius relations  are  also presented
in Figure \ref{figrm}. ( the black solid line (TM1) for pure \( npe \)
nuclear matter and the star with a crust (TM1 + Bonn + Negele + Vautherin
\cite{crust} - dotted line). The smaller sizes of  quarks star is
due to the fact that the pressure (Figure \ref{fesob}) reaches zero
for high densities (\( n_{B}^{m}\sim \, 0.26\, fm^{-3}=1.75\, n_{B}^{0} \))
then  to the symmetric nuclear matter (\( n_{B}^{0}\sim \, 0.15\, fm^{-3} \)).
On the surface such a star has higher density than the saturated
nuclear matter. This makes the star smaller and denser. The same situation
exists in the QMF approach when the bag constant is included. In the
NJL model the effective bag constant has dynamical origin (eq. \ref{beffe}).
Its main contributions come from quantum vacuum fluctuations and
scalar mesons. The strange quark \( s \) appears only for 
densities high enough (\( n_{B}>n^{s}_{B}=3.94\, n_{B}^{0} \)). For smaller densities
the quark star is built only from \( u \) and \( d \) quarks. It
is important to stress that even then the quantum vacuum fluctuation
come from all flavors including \( s \) quark and all antiparticles.
There are no \( s \) quarks for small densities in the strange quark
star only its quantum fluctuations. In the QMF approach there is no
quantum fluctuations at all. This is a significant difference between
the NJL and QMF approach. The another one is that the bag constant
in the QMF model must be added by hand. This make the QMF approach
unreliable for smaller densities (Figure \ref{fesob}). 

\begin{figure}
{\centering \resizebox*{15cm}{!}{\includegraphics{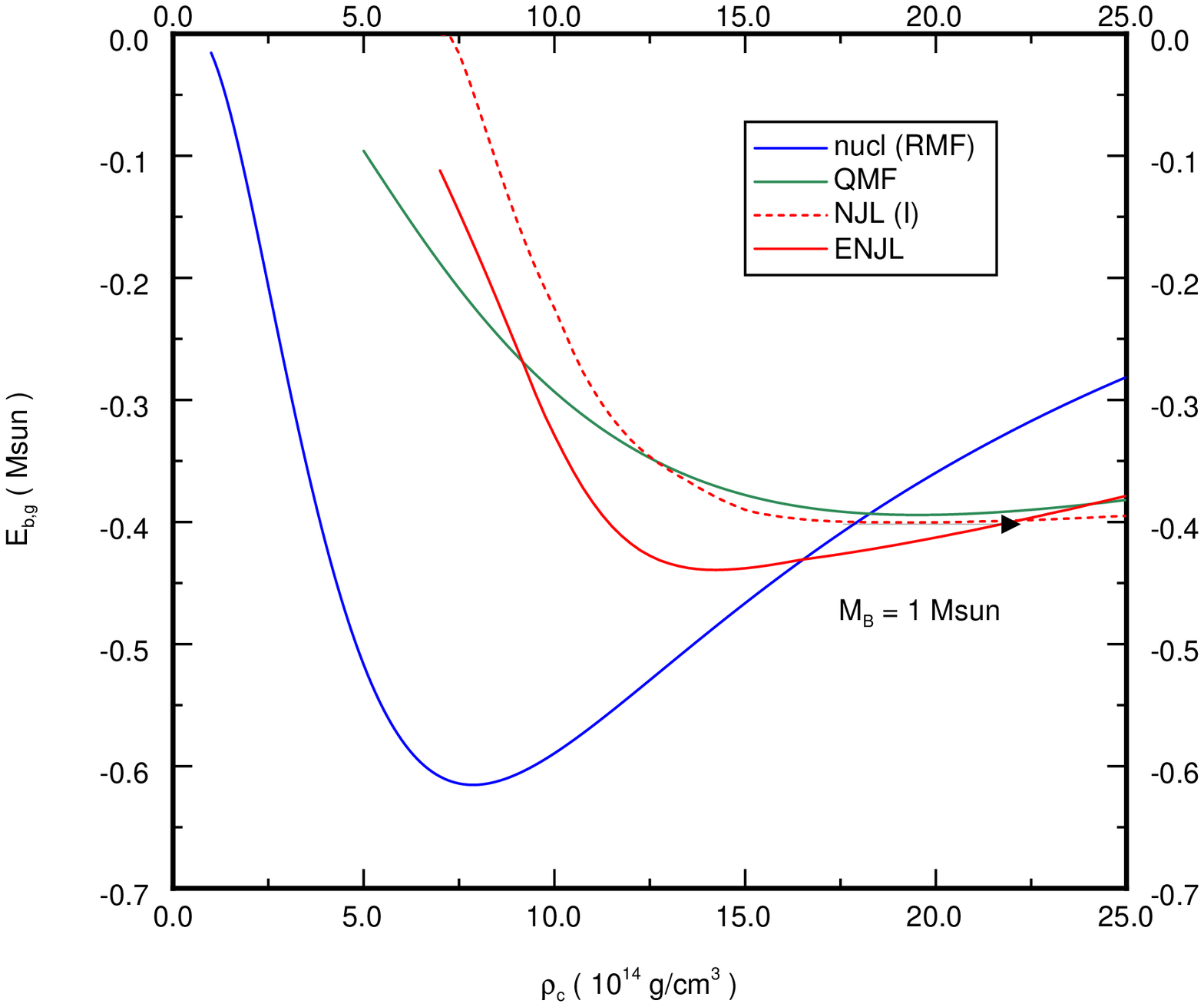}} \par}
\caption{\label{figgb}The gravitational binding energy of the quark strange
(ENJL and QMF approach) and neutron (RMF) star. The arrow shows a
possible transition from unstable neutron star to the strange one
with conservation of the baryon number \protect\( M_{B}=m_{n}c^{2}N_{B}\protect \).}
\end{figure}

The gravitational binding energy of a strange quark star  and QMF
approach) and neutron (RMF) star is presented in Figure \ref{figgb}.
The arrow shows a possible transition from unstable neutron star to
the strange one with conservation of the baryon number \( M_{B}=M_{odot},\, \, (M_{B}=m_{n}c^{2}N_{B}) \).

In the enlarged NJL model \cite{MCR} vector mesons are included.
There  contribution to the effective bag constant are positive (Figure
\ref{figb}). This makes that a  strange quark star possesses a bit bigger radius and mass  then
 in the NJL (I) model. 

\begin{figure}
{\centering \resizebox*{15cm}{15cm}{\includegraphics{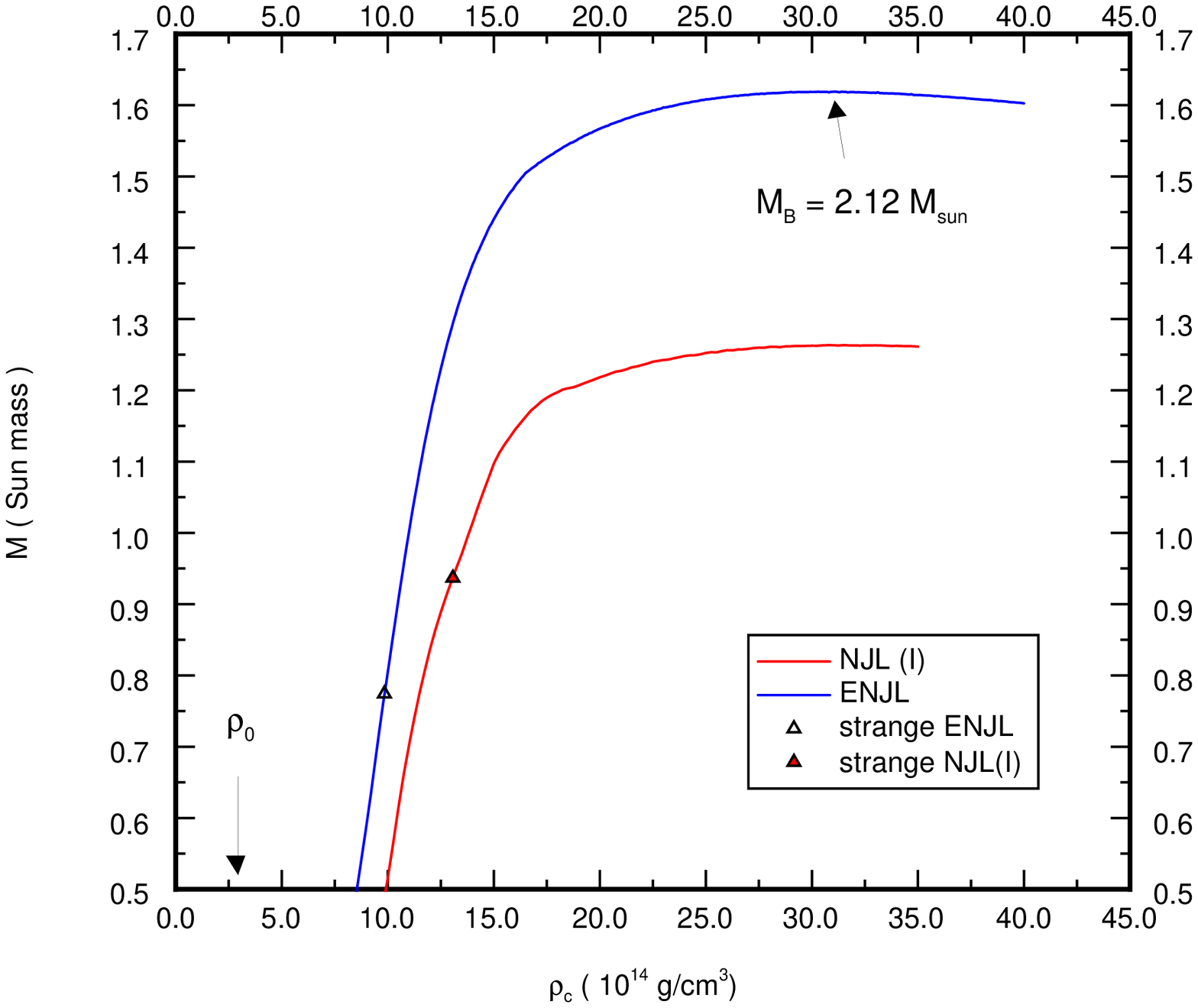}} \par}
\caption{The dependence for the quark star \protect\( M(\rho _{c})\protect \)
mass on the its central density \protect\( \rho _{c}\protect \).}
\label{figrhor}
\end{figure}

\begin{figure}
{\centering \resizebox*{15cm}{15cm}{\includegraphics{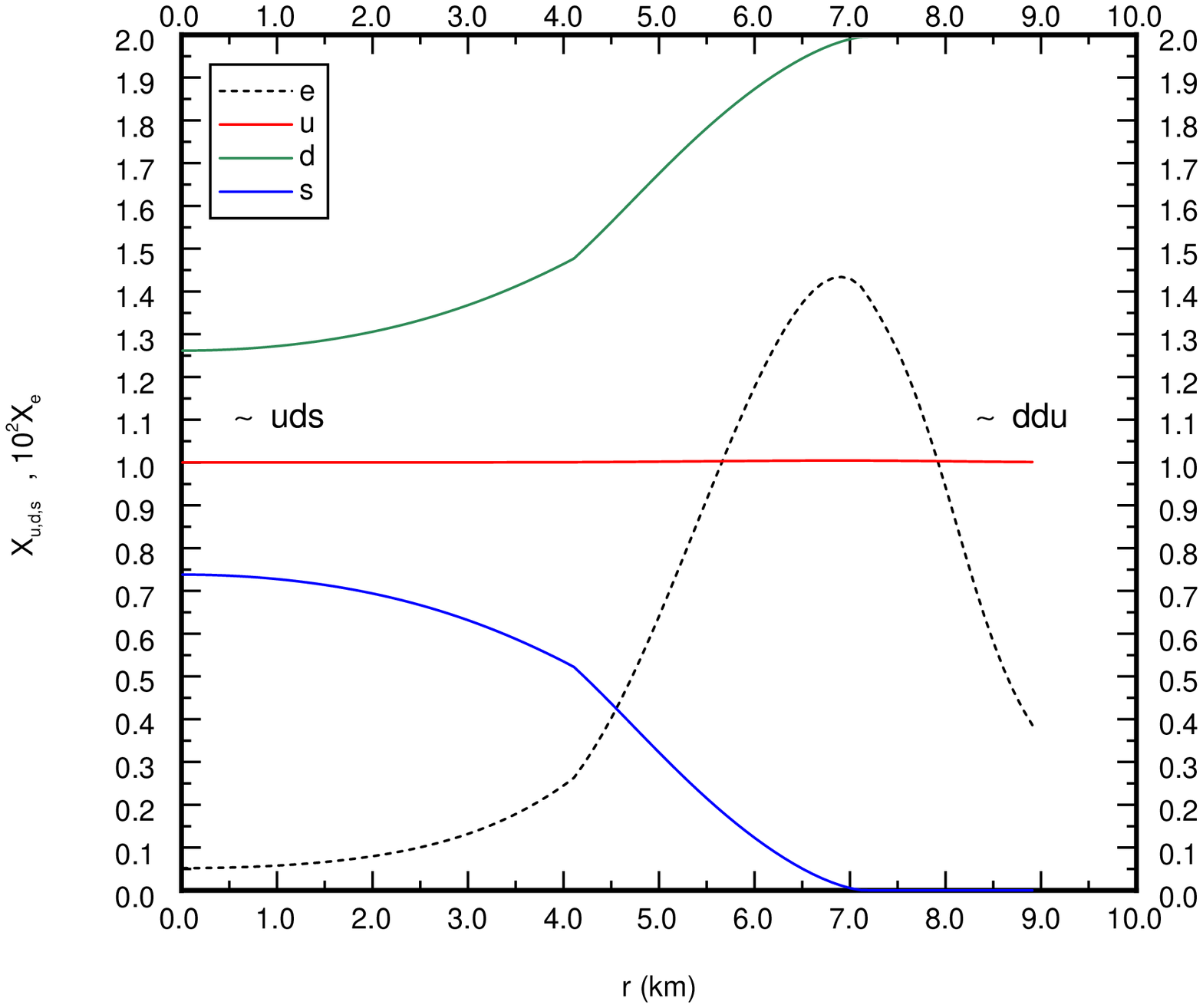}} \par}
\caption{The profile for effective quark Fermi momentum of the maximal quark
star (ENJL model) with the central density \protect\( \rho _{c}=3.11\, 10^{15}\, g/cm^{3}\protect \).}
\label{figmu}
\end{figure}

However, a maximum  stable strange quark star is obtained for  \( \rho _{2}=3.1\, 10^{15} \)
\( g/cm^{3} \) and has the following parameters \( M=1.61\, \, M_{\odot } \) and \( R=8.74\, \, km \).
The baryon number is this star is the same as in the case of pure neutron star with \( M_{B}=2.126\, M_{\odot } \).
Below the density  \( \rho _{s}=3.94\, \rho _{0}\, (\rho _{0}=2.5\, 10^{14}\, g/cm^{3}), \)
there are no strange quarks and  quark stars. Stable stars are
those with \( \frac{dM}{d\rho _{c}}>0 \) \cite{shap} (Figure \ref{figrhor}).
A gravitational binding energy for a strange quark is lower then
a neutron one for \( \rho >1.6\, 10^{15} \) \( g/cm^{3} \).
The \( M(\rho ) \) dependence for the quark star is presented in
Figure \ref{figrhor}. For the quark star with the maximal central
density \( \rho _{c}=3.11\, 10^{15}\, g/cm^{3} \) the star profile
is presented on the Figure \ref{figmu}. 
Quark and electron mass distributions inside the star are presented
in Figure \ref{figmu}. 
\begin{figure}
{\centering \resizebox*{15cm}{15cm}{\includegraphics{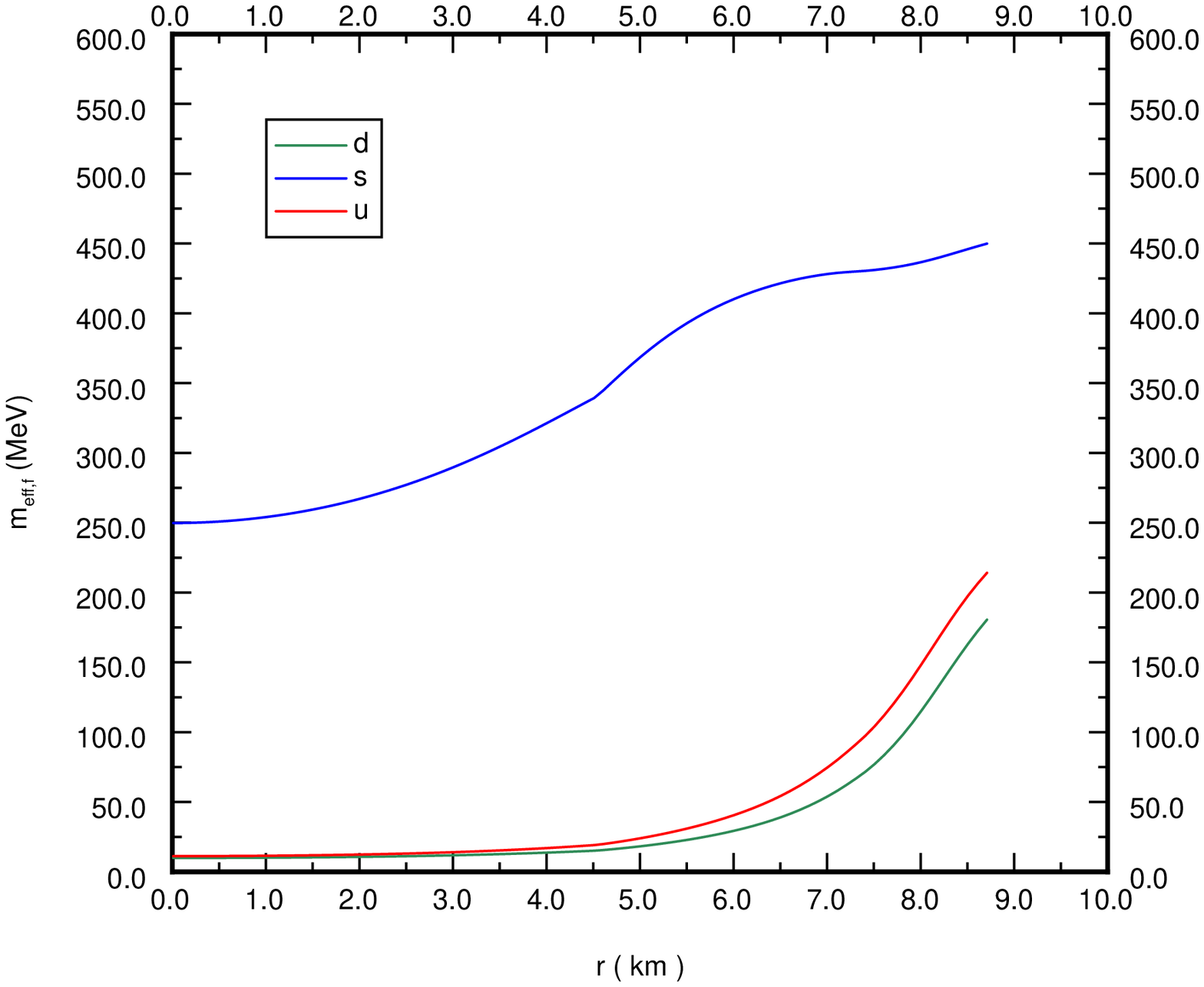}} \par}
\caption{The quark effective mass profile inside the maximal quark star (ENJL
model) with the central density \protect\( \rho _{c}=3.11\, 10^{15}\, g/cm^{3}\protect \).}
\label{Figdel}
\end{figure}
The quark partial fractions defined as \[
X_{f}=\frac{n_{f}}{(n_{u}+n_{d}+n_{s})}=\frac{3n_{f}}{n_{B}},\]
 where \( f=(u,d,s) \) are presented in  Figure \ref{figxs} .
 
\section*{\textcolor{blue}{Conclusion}}

The properties of the strange quark star in the bag model with \( B=B_{c} \)
and the current quark masses (\( m_{u}=m_{d}=0 \), \( m_{s}=150 \)~MeV)
are presented in \cite{GlenBook}. The star model base on the QMC 
is very similar (the dotted green line in  Figure \ref{figrm})
and close to the one for the ENJL model parametrized by the \( TM1 \) parameter
set (\( x_{v}=0.65 \), the solid violet curve \ref{figrm}). The
properties of  strange stars with quark masses  changing continuously
from constituent quark mass to the small current (see  Figure \ref{figd})
are presented in \cite{bomb1}. All these stars are more compact then
neutron stars (see the Figure \ref{figrm}) and are similar to those
of the NJL (I) model. 

In this paper the enlarged NJL model is used to construct the EoS
and properties of the strange quark star. The stable strange quark
star exists from minimal central density till the maximal one \( \rho _{2}=3.1\, 10^{15}\, g/cm^{3} \)
which gives the following star parameters \( M=1.61\, \, M_{\odot } \) and \( R=8.74\, \, km \). Its
baryon number is the same as for the pure neutron star with \( M_{B}=2.126\, M_{\odot } \).
The very similar strange star but less compact is obtained in the
solvable NJL model \cite{blaschke}. The gravitational binding energy
for a strange quark star is lower then the neutron one for densities \( \rho >1.6\, 10^{15} \)
\( g/cm^{3} \). The conversion of an unstable neutron star into a strange star is
an exciting subject which may help explain the gamma-ray bursts enigma
\cite{bomb2}. 

Similarly  to the QMF model the enlarged NJL one includes also the coupling to vector mesons. 
This is crucial for the quark star properties. Also the
quark \( s \) mass is important. The mass of a  quark  $s$ is also relevant because its smaller
mass causes that a strange quarks appear for lower densities. Nonzero strangeness of the matter
gives as a result a strange star.  
It is fascinating that the neutron and strange star properties are
strictly connected to inner structure of nuclei and nucleons. 
\newpage

\section*{\textcolor{blue}{References}}


\begin{thebibliography}{10}
\bibitem{latt}See for instance Karsch F., 
\href{http://xxx.lanl.gov/abs/hep-lat/9909006}{hep-lat/9909006}
\bibitem{NJL}Nambu Y. and  Jona-Lasinio G., Phys. Rev. 1961  \textbf{122} 345;
  \textbf{124} (1961) 246 
\bibitem{gast}Gastineau F., Aichelin J., 
  \emph{Strange baryons in the hot and dense medium within the Nambu - Jona - Lasinio model}, 
  \href{http://xxx.lanl.gov/abs/nucl-th/0201063}{nucl-th/0201063}
\bibitem{Vog91}Vogl U. and  Weise W., 1991 Prog. Part. Nucl. Phys.~\textbf{27}, 195 
\bibitem{Kle92}Klevansky S.P., 1992 Rev. Mod. Phys. \textbf{64}, 649 
\bibitem{Asa89}Asakawa M. and  Yazaki Y., 1989 Nucl. Phys. \textbf{A504}, 668 
\bibitem{Kli90a}Klimt S.,  Lutz M., and  Weise W., 1990 Phys. Lett. \textbf{B249}, 386 
\bibitem{Kle99}Schwarz T.M.,  Klevansky S.P. and Rapp G. 1999 
  Phys. Rev.~C \textbf{60}, 055205 
\bibitem{GlenBook}Glendenning N.K., 
  \emph{Compact Stars} (Springer-Verlag, 1997)
\bibitem{Weber}Weber F., 
  \emph{Pulsars as Astrophysical Laboratories for Nuclear and Particle Physics} (IoP, Bristol, 1999)
\bibitem{Hana00}Hanauske M. , Zschiesche  D. ,  Pal  S.,  Schramm  S., St\"{o}cker  H. and
  Greiner W. 2000 Astrophys. J. \textbf{537}, 958, 
  \href{http://xxx.lanl.gov/abs/astro-ph/9909052}{astro-ph/9909052}
\bibitem{Hana01} Hanauske M.,  Satarov  L.M.,  Mishustin I.N., 
  St\( \o  \)cker H.,  Greiner W., 2001 Phys.Rev. \textbf{D64}  043005, 
  \href{http://xxx.lanl.gov/abs/astro-ph/0101267}{astro-ph/0101267}
\bibitem{blaschke} Gocke C.,  Blaschke  D.,  Khalatyan  A., Grigorian  H. ,
  \emph{Equation of State for Strange Quark Matter in a Separable Model}, 
  \href{http://xxx.lanl.gov/abs/hep-ph/0104183}{hep-ph/0104183}
\bibitem{glen1} Glendenning N.K.,  Weber  F., 
  \emph{Possible evidence of quark matter in neutron star X-ray binaries},
  2001 Astrophys.J. 559  L119, \href{http://xxx.lanl.gov/abs/astro-ph/0003426}{astro-ph/0003426}
\bibitem{ABR01} Alford M.~G.,  Bowers  J.~A., and  Rajagopal K.
  2001 J. Phys. G \textbf{27}, 541.
\bibitem{biel}Schaffner--Bielich J. , 
\emph{The Role of Strangeness in Astrophysics - an Odyssey through Strange Phase}, \href{http://xxx.lanl.gov/abs/astro-ph/0112491}{astro-ph/0112491}
\bibitem{bw}Bodmer A. R., Phys. 1971 Rev.  
  \textbf{D4}, 1601; Witten E. 1984 Phys. Rev. D \textbf{30}, 272. 
\bibitem{Rel} Rehberg P.,  Klevansky S.P. and J. Hüfner J. 1996 Phys. Rev. \textbf{C53}  410. 
\bibitem{BO98}Buballa M. and Oertel M., 
  \href{http://xxx.lanl.gov/abs/hep-ph/9810529}{hep-ph/9810529};
  Buballa M. and Oertel M., 1998 Nucl. Phys. \textbf{A642}, 39; 1999 Phys. Lett. \textbf{B457}, 261 
\bibitem{thom} Bentz W., Thomas  A.W., 
  \emph{Stability of Nuclear Matter in the Nambu - Jona - Lasinio model,} 
  \href{http://xxx.lanl.gov/abs/nucl-th/0105022}{nucl-th/0105022}
\bibitem{riske}Scavenius O., M\'ocsy  A.,  Mishustin I.N.,  Rischke  
  D.H., 2001 Phys. Rev. \textbf{C64}, 045202, 
  \href{http://xxx.lanl.gov/abs/nucl-th/007030}{nucl-th/007030}
\bibitem{MCR}Ruivo M.C.,  Costa  P.,  de Sousa  C., 
  \emph{Pseudoscalar mesons in the asymmetric matter}, 
  \href{http://xxx.lanl.gov/abs/hep-ph/0109234}{hep-ph/0109234};
  Bernard V.,  Blin  A.H.,  Hiller  B.,  Meissner  V.-G.,  Ruivo M.C. 1993 
  Phys. Lett. B305  163-167, \href{http://xxx.lanl.gov/abs/hep-ph/0109234}{hep-ph/0109234}
\bibitem{fet}Fetter A.L.,  Walecka J.D.,
  \emph{Quantum Theory of Many-Particle Systems}, McGraw-Hill, New York, 1971 
\bibitem{torn}Th\( \o  \)rqvist N.A., 
  \emph{The linear U3xU3 Sigma Model, the sigma resonance and the spontaneous breaking of symmetry}, 
  \href{http://xxx.lanl.gov/abs/hep-ph/9711483}{hep-ph/9711483}
\bibitem{rm}Feynman R., 
  \emph{Statistical Mechanics}, W.A.Benjamin, Inc. Reading 1972.\\
  Ma\'{n}ka R.,  Kuczy\'{n}ski J.,  Vittiello  G.V. 1986 Nucl. Phys. \textbf{B276}, 533; 
  Annals of Physics, \textbf{199}, Ma\'{n}ka R.,  Vittiello G.V., \textbf{199}, (1990), 61. 
\bibitem{rm1}Ma\'{n}ka R., Bednarek I. and Przyby\l{}a 2000 Phys. Rev.\textbf{C62} 015802; 
  \href{http://xxx.lanl.gov/abs/nucl-th/0001017}{nucl-th/0001017}
\bibitem{FarhiJaffe84}Farhi E. and Jaffe R.L., 1984 Phys.~Rev.~\textbf{D30}  2379
\bibitem{MITbag}Chodos A., Jaffe R.L.~, Johnson  K.~, Thorn C.B.~ 
and Weisskopf V.F.~, 1974 Phys. Rev.~\textbf{D9}  3471; \\
Chodos  A.~, Jaffe R.L.~, Johnson K.~, and Thorn C.B.~ 1974 Phys.~Rev.~\textbf{D10}
 2599
\bibitem{adami} Adami C. and  Brown G. E. 1993 Phys. Rep. \textbf{234}, 1; 
Xue-min Jin and  Jennings B.K. 1997 Phys. Rev. \textbf{C55}, 1567; 
\bibitem{aust} Guichon P. 1988 Phys. Lett. \textbf{B200}, 235; 
 Guichon P.,   Saito  K.,  Rodionov E.,  Thomas  A.W., 
 \emph{The role of nucleon structure in finite nuclei}, 
 \href{http://xxx.lanl.gov/abs/nucl-th/9509034}{nucl-th/9509034};
 Shen H., Toki H. , 
  \emph{Quark mean field model for nuclear matter and finite nuclei}, 
  \href{http://xxx.lanl.gov/abs/nucl-th/9911046}{nucl-th/9911046} 
\bibitem{toki1} Shen H., Toki H.  1978 Phys. Rev. \textbf{D18}, 4187,  
  \emph{Quark mean field model for nuclear matter and finite nuclei}, 
  \href{http://xxx.lanl.gov/abs/nucl-th/9911046}{nucl-th/9911046} 
\bibitem{wal}Serot B.D., Walecka J.D. 
  \emph{Recent Progress in Quantum Hadrodynamics}, 1997 Int. J. Mod. Phys. 
  \textbf{E6}, 515-631, 
  \href{http://xxx.lanl.gov/abs/nucl-th/9701058}{nucl-th/9701058}
\bibitem{toki}Sumiyoshi K., Toki  H. 1994 ApJ, \textbf{422}, 700 
\bibitem{rm2}Ma\'{n}ka R., Bednarek I. Int. J. Mod. Phys. D10 (2001) 607-624,
  \href{http://xxx.lanl.gov/abs/nucl-th/0012026}{nucl-th/0012026}
\bibitem{kub4} Kubis S.,  Kutschera  M. 1997 Phys. Lett. \textbf{B399}, 191
\bibitem{liu3} Liu B., Greco V. , Baran V., Colonna M. , Di Toro  M., 
  \emph{Asymmetric nuclear matter: the role of the isovector scalar channel}, 
  \href{http://xxx.lanl.gov/abs/nucl-th/0112034}{nucl-th/0112034}
\bibitem{Bogut77}Boguta J and Bodmer A R 1977 \textit{Nucl. Phys.} \textbf{A292} 413 
\bibitem{shap}Shapiro S L and Teukolsky S A, 1983 \textit{Black hols, white dwarfs
and neutron stars} New York 
\bibitem{tm1} Sugahara  Y. and  Toki H. 1994 Prog. Theo. Phys. \textbf{92},  803
  Heiselberg H., \emph{Neutron star: recent developments}, 
  \href{http://xxx.lanl.gov/abs/nucl-th/9912002}{nucl-th/9912002};
  Heiselberg H.,  Hjorth-Jensen M., 
  \emph{Phases of Dense Matter in Neutron Stars,} 1999 
\bibitem{blaschke1}Blaschke D., Burau  G.,  Volkov M.K., Yudichev V.L. 2001
   Eur. Phys. J. \textbf{A1}1, 319, 
   \href{http://xxx.lanl.gov/abs/hep-ph/0107126}{hep-ph/0107126}
\bibitem{tov}Tolman R.C., Phys. Rev. 1939 \textbf{55}  364;\\
   Oppenheimer J.R. and  Volkoff G.M. 1930 Phys.Rev. \textbf{55}  374
\bibitem{crust} Negele J.W.,  Vautherin D. 1973 Nucl. Phys. \textbf{A207}, 298;
   Haensel P.,  Pichon  B. 1994 Astron. Astrophys. \textbf{283}, 313.
\bibitem{bomb1}Bombaci I.,  Datta A.V. B. 2000 Astrophys.J. 541  L71; 
   \emph{Rapidly rotating strange stars for a new equation of state of strange quark matter},
   \href{http://xxx.lanl.gov/abs/astro-ph/0009328}{astro-ph/0009328};
 Dey M.,  Bombaci I.,  Dey J., Subharthi Ray,  Samanta B. C.,
   \emph{Strange Stars with Realistic Quark Vector Interaction and Phenomenological Density-dependent Scalar Potential}, 
   1998 Phys.Lett. \textbf{B438} 123-128; 
   1999 Addendum-ibid. \textbf{B447}  352-353, 
   \href{http://xxx.lanl.gov/abs/astro-ph/9810065}{astro-ph/9810065}
\bibitem{bomb2} Li X.-D., Bombaci I. ,  Dey M.,  Dey J.,  van den Heuvel E. P. J.,
    1999  Phys.Rev.Lett. \textbf{83}  3776-3779, 
    \href{http://xxx.lanl.gov/abs/hep-ph/9905356}{hep-ph/9905356};
     Bombaci I., 
\emph{Strange star candidates}, 
\href{http://xxx.lanl.gov/abs/astro-ph/0201369}{astro-ph/0201369}
\end{thebibliography}
\end{document}